# Lossy phononic metamaterials for valley manipulation

Shunda Yin[1†], Qiuyan Zhou[1†], Yuxiang Xi[1], Weiyin Deng[1*], Wei Chen[2,3*], Jiuyang Lu[1*], Manzhu Ke[1], and Zhengyou Liu[1,4*]

[1]Key Laboratory of Artificial Micro- and Nanostructures of Ministry of Education and School of Physics and Technology, Wuhan University, Wuhan 430072, China
[2]National Laboratory of Solid State Microstructures, School of Physics, and Collaborative Innovation Center of Advanced Microstructures, Nanjing University, Nanjing 210093, China
[3]Jiangsu Physical Science Research Center and Jiangsu Key Laboratory of Quantum Information Science and Technology, Nanjing University, Nanjing 210093, China
[4]Institute for Advanced Studies, Wuhan University, Wuhan 430072, China

†These authors contributed equally to this work.
*Corresponding author.
Emails: dengwy@whu.edu.cn; pchenweis@gmail.com; jylu@whu.edu.cn; zyliu@whu.edu.cn

Non-Hermitian physics characterized by complex band spectra has established a new paradigm in condensed matter systems and metamaterials. Recently, non-Hermitian gain and nonreciprocity are deliberately introduced to valley manipulation, leading to various phenomena beyond the Hermitian scenarios, such as the amplified topological whispering gallery modes as an acoustic laser. In contrast, pure loss is inevitable in practice and generally regarded as a detrimental factor. Here, we reveal that the coupling loss can manipulate valley degrees of freedom in a phononic metamaterial. Three distinct valley-related effects, including valley-resolved nonreciprocity that functions as a valley filter, valley-dependent skin effects where bulk states from different valleys localize at opposite boundaries, and valley-projected edge states with boundary-dependent lifetimes that leads to an anomalous beam splitting, are demonstrated through theoretical analysis and airborne sound experiments. Owing to the easy preparation of loss, our findings shed light on both non-Hermitian and valley physics and may facilitate innovative applications of valley-related devices.

In recent years, non-Hermitian (NH) physics, which explores the unique properties and novel phenomena of open systems described by NH Hamiltonians, has attracted extensive interest in condensed matter systems and metamaterials [1-3]. Under NH interactions of gain, loss or nonreciprocity, the eigenvalues of the NH Hamiltonian are generally complex numbers, with their real and imaginary parts corresponding to energy storage and input/output. Meanwhile, the NH eigenstates are nonorthogonal and can even coalesce with each other [4]. Regarding topological properties, in addition to the wave-function topology that predicts conventional boundary states [5,6], complex eigenvalues introduce a new type of topology, i.e., spectral topology [7,8]. The spectral topology is characterized by the winding number of the complex eigenvalues, which guarantees the NH skin effect, that is, allowing bulk states to manifest as skin modes collapsing to open boundaries [9-11]. So far, various configurations of skin effects have been experimentally visualized on different platforms, including one-dimensional (1D) skin effect [12-15], higher-order skin effect [16,17], geometry-dependent skin effect [18-21], and hybrid skin-topological effect [22-25]. This not only greatly promotes the development of the NH physics, but also facilitates advances in applications [26-28], including topological switch and sensor.

Discrete valleys, as a versatile degree of freedom, have been exploited as efficient carriers for processing and storing information and energy [29,30]. A variety of valley-related phenomena, such as valley Hall effect [31-34], valley filters [35-38], and valley Zeeman effect [39-42], have been explored both theoretically and experimentally. Moreover, bulk valley transport with locked chirality [43-45] and topological valley edge transport with immunity to lattice imperfections [47-52] have been extensively reported in optical, acoustic, and on-chip metamaterials. Recently, the introduction of the non-Hermiticity has opened new avenues for valley manipulations [53-56]. For instance, a proposed NH Moiré valley filter can be driven by a valley-resolved NH skin effect with the valley-resolved nonreciprocal transport [54]. In practice, with gain medium tuned by the electro-thermoacoustic coupling, amplified topological whispering gallery modes can be engineered as acoustic lasers in phononic metamaterials [55]. And NH Dirac cones with valley-dependent lifetimes enabled by nonreciprocal coupling with operational amplifier have been realized in electric circuits [56]. However, the gain or nonreciprocity interactions are typically implemented by active devices, which must be deliberately designed and complicate the structure. In contrast, pure loss naturally occurs in materials and is easy to implement, yet it is generally regarded as a detrimental factor. It is therefore desirable to explore the beneficial effects of loss and its potential applications in valley physics.

In this work, we reveal three novel NH valley phenomena in a honeycomb lossy phononic metamaterial. The coupling loss is readily implemented by drilling holes on the coupling tubes and sealing them with the sound-absorbing sponges. Remarkably, the complex bulk dispersions exhibit valley-resolved nonreciprocity, therefore functioning as a NH valley filter. Meanwhile, such metamaterial possesses nontrivial spectral topology and the ensuing the valley-dependent skin effect, where the bulk states near different valleys are localized at the opposite boundaries. And it hosts wave-function topology simultaneously, leading to NH valley-projected edge states with boundary-dependent lifetimes, which display an anomalous beam splitting beyond the Hermitian scenario. These unique NH valley phenomena are corroborated by airborne sound experiments, which are in excellent agreement with the theoretical predictions and numerical simulations.

**Tight-binding model.**

We start from a honeycomb lattice with coupling loss and staggered on-site energy, as depicted in Fig.1a. The corresponding Hamiltonian in momentum space reads

$$H = (d_x - i\gamma)\sigma_x + d_y\sigma_y + d_z\sigma_z, \tag{1}$$

where $d_x = t_0 + 2t_0 \cos\frac{\sqrt{3}k_y}{2}\cos\frac{k_x}{2}$, $d_y = -2t_0 \sin\frac{\sqrt{3}k_y}{2}\cos\frac{k_x}{2}$, $d_z = m$, $\sigma_i$ ($i = x, y, z$) are Pauli matrixes. Here, $\mathbf{k} = (k_x, k_y)$ is the Bloch wavevector, $m$ is on-site energy, $t_0 - i\gamma$ is the intracell NH hopping with $\gamma$ marks the strength of loss, while $t_0$ is the intercell Hermitian hopping. The lattice constant is set to unity for simplicity. Under the coupling loss, the system possesses the spinless anomalous time-reversal symmetry (aTRS), expressed as $U_T H^t(\mathbf{k})U_T^{-1} = H(-\mathbf{k})$, where $U_T$ is a unitary matrix and the superscript $t$ denotes transpose operation. Note that, $H^t(\boldsymbol{k})$ and $H(\boldsymbol{k})$ have the same eigenvalues, giving rise to energy spectra $E(\boldsymbol{k}) = E(-\boldsymbol{k})$ (Supplementary S-I A).

The complex bulk band dispersions near the $K$ and $K'$ valleys are shown in Fig. 1b, with band gaps opened. Interestingly, as the real and imaginary parts of the dispersions are shown, the complex bulk band dispersions reveal the valley-resolved nonreciprocity, arising jointly from aTRS and the imaginary energy reversal properties (Supplementary S-I B). Physically, the imaginary part of the eigenvalues determines the lifetimes of the states: the smaller the imaginary eigenvalues, the longer the lifetimes. The left-moving states near the $K$ valley possess longer lifetimes than the right-moving states within the same valley, while the behavior is reversed at the $K'$ valley as required by aTRS. That is, our model can only support the right-moving states near the $K'$ valley and the left-moving states near the $K$ valley in long-time limit, thus exhibiting a valley-resolved nonreciprocal transport, as depicted by the inset in Fig. 1b. Based on this property, one can realize unidirectional transport of the bulk valley states and design valley filters.

Next, we analyze the topological properties of the system, including both the spectral topology and the wave-function topology. When examining the spectral topology of the system, the opposite spectral winding numbers of $K$ and $K'$ valleys along some special directions give rise to valley-dependent skin effects. Figure 1c presents the spectral winding numbers along the $k_{y'}$ direction as a function of $k_{x'}$. It shows that the spectral winding numbers are inverted near the $K$ and $K'$ valleys, indicating that the states near different valleys will be localized at the opposite boundaries along the $x'$ direction, constituting the so-called valley-dependent skin effect. To verify this, we consider a ribbon with periodic boundary conditions in the $x'$ direction and open boundary conditions in the $y'$ direction, and calculate the projected dispersions. The corresponding results are shown in Fig. 1d, where the positive (negative) values of the colors indicate the localization degree of the bulk states at the upper-left (lower-right) boundary. Clearly, the states near the $K$ and $K'$ valleys are located at the opposite boundaries, exhibiting a valley-dependent feature. This can be further visualized by the eigenfield distributions of the states near the $K$ and $K'$ valleys shown in the inset of Fig. 1d, which are confined at the upper-left and lower-right boundaries, respectively, demonstrating the valley-dependent skin effect. See detailed calculations in Supplementary S-I C.

The wave-function topology of the system can be characterized by the valley Chern number, which guarantees the appearance of the valley-projected edge states. Figure 1e shows the phase diagram described by the valley Chern number $C_K$ as a function of $\gamma/t_0$ and $m/t_0$ ($C_{K'} = -C_K$, as required by aTRS), where there exist three topologically distinct phases. In the orange and purple regions (referred to as phases I and II, respectively), the system has a bulk band gap with $C_K = \frac{1}{2}$ and $-\frac{1}{2}$, respectively; at the gray region, the system has no bulk band gap and the valley Chern number cannot be well defined. According to the bulk-boundary correspondence, the valley-projected edge states will emerge at the interface between two phases with opposite valley Chern number. Therefore, when phases I and II are combined to form an I-II-I heterostructure (Supplementary S-I D), the valley-projected edge states appear at both interfaces I-II and II-I, as shown by the complex projected dispersions in Fig. 1f. Notably, due to inheriting the lifetimes of the bulk, only the valley-projected edge states at the interface I-II persist in the long-time limit, whereas those at the interface II-I decay rapidly, exhibiting a boundary-selective behavior.

**Acoustic NH valley filter.**

The tight-binding model can be directly implemented in a phononic metamaterial of cavity-tube structure filled with air, as illustrated in the enlarged view inserted in Fig. 2a. The schematic of the acoustic unit cell of the NH honeycomb lattice is depicted in Fig. 2b, corresponding to the yellow rhombus region in the inset in Fig. 2a. Structurally, the unit cell is composed of two nonequivalent acoustic cavities (yellow), labeled as 1 and 2, which are coupled through some identical rectangular waveguides (brown). The acoustic cavities have a side length of $w_1 = 6$ mm, and heights of $h_1 = 30.4$ mm and $h_2 = 29.6$ mm. The width of the waveguides is $w_2 = 5$ mm and the lattice constant is $a = 40$ mm. Green circles with diameters of $d = 3.6$ mm on the waveguide represent three holes, which are used to introduce loss by sealing with the sound-absorbing sponges in the experiments. In simulations, the designed loss is realized by adding an imaginary part ($30i$ m/s) on the sound velocity of the corresponding waveguides (Supplementary S-II). Figure 2c presents the simulated complex bulk band dispersions near the $K$ valley for the acoustic unit cell. The complex projected dispersions of the bulk states are displayed in Fig. 2d. One can see that the left-moving states near the $K$ valley and the right-moving states near the $K'$ valley possess longer lifetimes and thus exhibit a valley-resolved nonreciprocity, consistent with the theoretical results.

Based on the valley-resolved nonreciprocity, we realize unidirectional transport of bulk valley states and design an acoustic valley filter. Figure 2a shows the designed acoustic valley filter, which consists of two types of domains A and B (separated by the blue dashed lines), corresponding to acoustic Hermitian and NH honeycomb lattices, respectively. The Hermitian domain is constructed similarly to the NH domain but without holes on their waveguides, and it hosts two valleys with real eigenfrequencies at the $K$ and $K'$ points. Such a sandwich-shaped structure (denoted as ABA) is a valley filter. To show this, ten point-sources are placed on the left boundary of the sample to excite the acoustic pressure field, and the sound pressure and phase were measured by drilling holes on the top of the entire sample, which are sealed when not in use. The simulated pressure field distribution of ABA at 5.86 kHz is shown in Fig. 2e. In experiment, we extract the sound signals within the green boxes at the input and output ports of ABA, and perform a 2D Fourier transform. The corresponding results are respectively shown in the upper panels of Fig. 2f. As we can see, the hybrid valley states containing both $K$ and $K'$ valleys are excited at the input port, while only the states near the $K'$ valley arrive at the output port. The experimental data well capture the simulations in the lower panels of Fig. 2f, demonstrating the valley filtering effect. It is worth noting that, compared to the previous proposals [38,54], the current scheme utilizes a NH strategy and has the advantages of a simple structure and no need to modify the system boundaries.

**Valley-dependent skin effect.**

In the following, we numerically and experimentally validate the valley-dependent skin effect. Figure 3a shows the simulated real part of the projected dispersions of the ribbon, which is periodic in the $x'$ direction and finite in the $y'$ direction. The positive (negative) values of the colors represent that the states are localized at the upper-left (lower-right) boundary, and the eigenfield distributions of the states denoted by stars are displayed in the inset. It shows that the states near the $K$ and $K'$ valleys are confined at the opposite boundaries, exhibiting the valley-dependent property and in good agreement with the theoretical results.

Such valley-dependent skin effect can be clearly revealed by our airborne sound experiments. We construct a finite-size phononic metamaterial with open boundaries in the $y'$ direction and radiative boundaries in the $x'$ direction, and measure the pressure field distribution for fixed frequencies. The source excites at each cavity, and a microphone probes the pressure response at the cavity at the next period along the $x'$ direction. Figure 3b shows the measured pressure field distribution at 5.86 kHz, which is strongly confined at the upper-left and lower-right boundaries, confirming the skin states in the finite-size sample. We further verify that these skin states are valley-dependent. We position a point source at the center close to the lower-right boundary to excite skin states there (Supplementary S-III). The measured projected band dispersions are displayed visually with the bright colors in Fig. 3c, the experimental data agrees well with the simulation denoted by white dots, and shows that the skin states confined at the lower-right boundary are mostly located near the $K'$ valley. Similarly, the point source positioned close to the upper-left boundary can excite the skin states there well, and the measured projected dispersions show that the skin states localized at the upper-left boundary are mostly located near the $K$ valley, as shown in Fig. 3d. The simulations together with experiments demonstrate the valley-dependent feature of the skin effect.

**Valley-projected edge states with boundary-dependent lifetimes.**

We then study the valley-projected edge states with boundary-dependent lifetimes. Figure 4a provides the phase diagram revealed by the real part of the band-edge frequencies at $K$ point versus the height difference of the two nonequivalent cavities of the acoustic unit cell. Evidently, the band gap remains closed when $\Delta h$ is small (here $|\Delta h| < 0.4$ mm), while opens and forms valleys as $|\Delta h|$ gradually increases ($|\Delta h| > 0.4$ mm). We have checked that the valleys for $\Delta h < -0.4$ mm and $\Delta h > 0.4$ mm possess opposite valley Chern numbers, i.e., $C_K = \frac{1}{2}$ for the former and $C_K = -\frac{1}{2}$ for the latter, referred to as phases I and II, respectively. Figure 4b displays the bulk band dispersions (with band gap of 5.51-6.00 kHz) of phase I with $\Delta h = -3$ mm, and its mirror counterpart (phase II) with $\Delta h = 3$ mm has exactly the same bulk band dispersions but the inversed valley topology. Similarly, we consider a ribbon of heterostructure I-II-I composed of phases I and II and calculate its complex projected dispersions, as shown in Fig. 4c. Evidently, owing to their inheritance of the lifetime profile of the bulk states, the valley-projected edge states (blue) along the interface I-II possess longer lifetimes than those (red) along the interface II-I, indicating their boundary dependence that is consistent with the tight-binding results.

The valley-projected edge states with boundary-dependent lifetimes can be demonstrated by the simulations and experiments. As shown in Fig. 4d, we fabricate two samples of interfaces I-II and II-I, and place a point source at the center of the interfaces to excite the valley-projected edge states. As exemplified by the pressure field distributions simulated at 5.75 kHz, the valley-projected edge states on interface I-II propagate along $\pm x$ directions with negligible attenuation, while those on interface II-I scarcely propagate. In experiments, we measure the projected dispersions for the two cases by Fourier transforming the measured pressure field at the interfaces, respectively. As displayed in Figs. 4e and 4f, the experimental data (color map) precisely captures the simulation results (white dots) and shows the good excitation of both the right- and left-moving valley-projected edge states on interface I-II, but no excitation of that on interface II-I. All these simulation and experiments explicitly demonstrate boundary dependence of the edge states.

Such a property can lead to an anomalous beam splitting at the intersection. As presented in Fig. 4g, we precisely fabricate an arrow-shaped interface intersection, which contains four ports marked by U, L, D, and R. A point source is positioned at the port U to excite the valley-projected edge states, and the corresponding simulated pressure field distribution is presented by color map in Fig. 4g. It shows that the $K$ valley-projected edge states can only transport along the interface I-II and reach the output port L, but not ports D and R. This behavior is significantly distinct from the Hermitian case, in which the $K$ valley-projected edge states can reach both ports L and R. Experimentally, we measure the transmitted pressure at the ports L, D and R. The results are presented in Fig. 4h, which shows that port L allows pressure transmit, while ports D and R forbid it, in agreement with the simulations.

**Discussion.**

In summary, we have observed three unique NH valley-related phenomena of both the bulk and edge states in a lossy phononic metamaterial. Benefiting from the valley-resolved nonreciprocity of the complex energy spectra, we have designed a NH valley filter and demonstrated unidirectional transport of bulk valley states. Unlike previous proposals that require complicated structural designs or boundary modifications, our design provides a simple and universal approach. Based on the spectral topology, our work has built an experimental bridge between valley and skin effects, and demonstrated the valley-dependent skin effects. At the boundary, the NH valley-projected edge states, protected by the wave-function topology, are verified to possess boundary-dependent lifetimes and can be exploited to achieve anomalous beam splitting. Our work reveals a fascinating interplay between non-Hermiticity and valley physics, and paves the way for applications of NH valley-related devices.

**Methods**

**Numerical simulation.** All the simulations are performed by the commercial COMSOL Multiphysics solver package. The systems are filled with air (with sound velocity of $343 \text{ m/s}$ and air density of $1.29 \text{ kg/m}^2$ at room temperature). We focus on the dipole mode of the cavity along the $z$ direction. The resin used for 3D printing is modeled as acoustic rigid wall, considering its huge impedance compared to the air. The simulated band dispersions and pressure field distributions of ribbon in the main text are calculated by applying Bloch boundary conditions in the periodic directions in corresponding unit cell or supercells. For brevity, trivial edge states localized at the edges, rather than the interfaces, are eliminated in all projected dispersions for both tight-binding calculations and acoustic simulations. All the simulated pressure field distributions are normalized by the corresponding maximum values. The spatial Fourier spectra in the lower panel of Fig. 2f are obtained by Fourier transforming the corresponding simulated acoustic pressure fields. In Fig. 3a, the specific value of the color is calculated by $D = \sum_{x \in L_1} |\psi(x)|^2 - \sum_{x \in L_2} |\psi(x)|^2$, where $\psi(x)$ is normalized eigenstate, $L_1$ and $L_2$ are the defined boundary lengths for upper-left and lower-right boundary, respectively. Specifically, the ribbon used for the calculation contains $48$ cavities, and each of $L_1$ and $L_2$ contains the $8$ cavities closest to the corresponding boundary.

**Experimental measurement.** The experimental sample is fabricated by 3D printing techniques of resin material. The geometric tolerance is $\sim 0.1 \text{ mm}$. The thicknesses of the cavities and tubes are set as $2 \text{ mm}$. Circular holes with diameters of $3.7 \text{ mm}$ are perforated on each cavity for the insertion of loudspeakers (with a diameter of $1.4 \text{ mm}$, can be viewed as point sound sources) or microphones (B&K Type 4183) to excite or probe the pressure field distributions of the crystal. The amplitude and phase of the acoustic field are recorded and frequency-resolved by a multi-analyzer system (B&K Type 3560B). The spatial Fourier spectra in the upper panel of Fig. 2f, and the projected dispersions in Figs. 3c, 3d, 4e and 4f are obtained by Fourier transforming the corresponding measured acoustic pressure fields. In Fig. 3b, the pressure field distribution is normalized by the maximum value.

**Data availability**

All the data supporting this study are displayed in the main text and Supplementary Information. Additional data related to this study are available from the corresponding authors upon request.

**Code availability**

The codes that support the results of this study are available from the corresponding authors upon request.

**Acknowledgements**

This work is supported by the National Key R&D Program of China (Nos. 2022YFA1404900, 2022YFA1404500, 2023YFB2804701, 2022YFA1204701), National Natural Science Foundation of China (Nos. 12222405, 12374409, 12504519, 12574024, 12574484, 12574051, 12222406), Natural Science Foundation of Hubei Province of China (No. 2024AFB712), the Natural Science Foundation of Jiangsu Province (Nos. BK20250008, BK20233001), and the Fundamental Research Funds for the Central Universities (Nos. KG202501 and 2024300415).

**Author contributions**

All authors contributed extensively to the work presented in this paper.

**Competing financial interests**

The authors declare no competing financial interests.

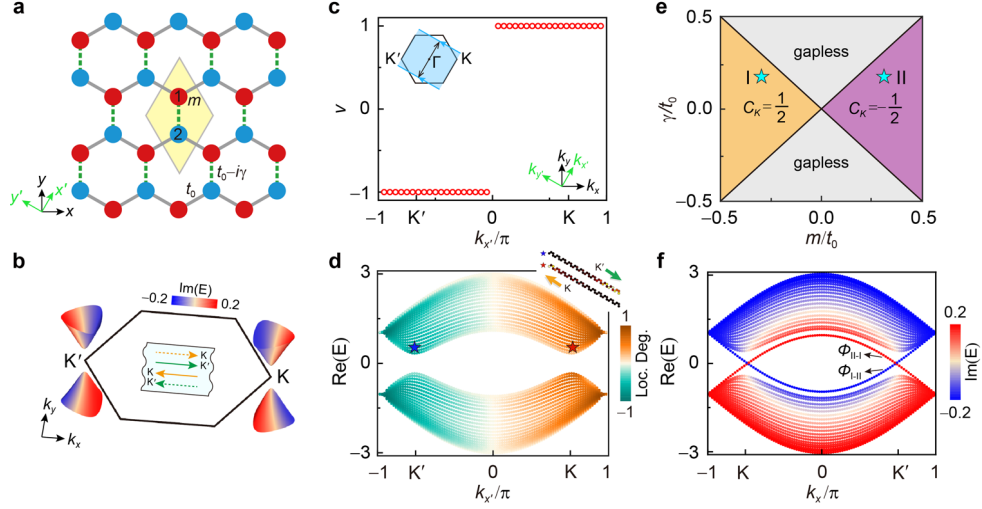

**Fig. 1 | NH valley phenomena in a lossy honeycomb lattice. a**, Schematic of the NH lattice model. The yellow area denotes the unit cell, which consists of two nonequivalent sites 1 and 2 with opposite on-site energies. **b**, Complex bulk band dispersions around the $K$ and $K'$ valleys. The colors denote the imaginary part of the dispersions. The black lines mark the first BZ. Inset: Schematic of the valley-resolved nonreciprocal transport. **c**, Spectral winding numbers along the $k_{y'}$ direction varies with $k_{x'}$. Inset: Schematic of the parameter evolution path. **d**, Real part of the projected dispersions of a ribbon in the inset. The positive and negative values of the colors indicate the localization degree of the bulk states at the upper-left and lower-right boundaries of the ribbon, respectively. Inset: Field distributions of two eigenstates denoted by blue and red stars in the dispersions. **e**, Phase diagram determined by the valley Chern number $C_K$ in the $\gamma/t_0$ and $m/t_0$ plane. The gray regions represent bandgap closure. The cyan stars denote the phases with the specific parameters used in **f**. **f**, Complex projected dispersions of a ribbon with periodicity in the $x$ direction, containing interfaces I-II and II-I. The colors denote the imaginary part of the dispersions. The parameters are chosen as $t_0 = 1$, $\gamma = 0.2$, and $m = -0.3$.

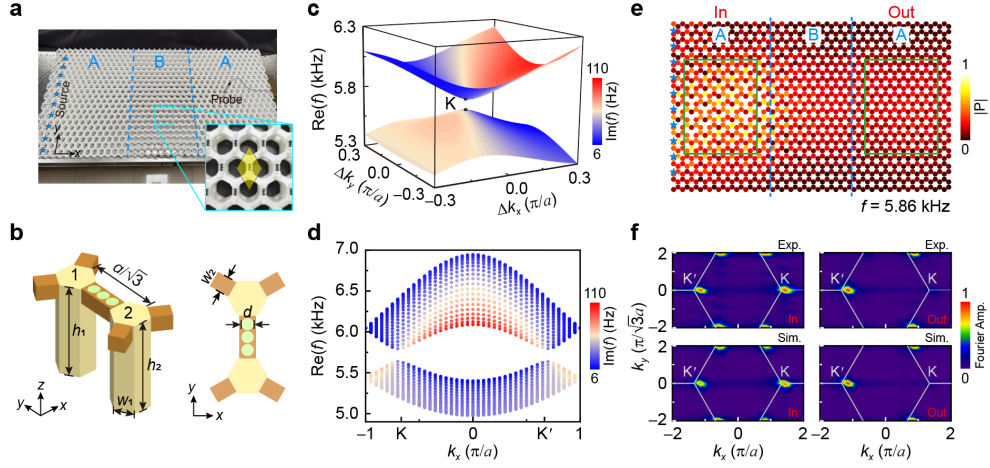

**Fig. 2 | Realization of the acoustic NH valley filter. a**, Photograph of the sample. Inset: Enlarged view of the NH domain $B$, where the yellow shadow marks a unit cell. The designed loss is achieved by the holes on the waveguides sealed with the sound-absorbing sponges. **b**, Schematic of the acoustic unit cell. Green circles represent the holes on the waveguide. **c**, Complex bulk band dispersions near the $K$ valley. **d**, Projected dispersions of a ribbon with periodic boundary conditions in the $x$ direction. **e**, Simulated pressure field distribution at 5.86 kHz, which is excited by ten point-sources (blue stars) on the left of the $ABA$ sandwich-shaped structure. **f**, Top panel: Measured spatial Fourier spectra of the acoustic field within the green rectangles at the input (left) and output (right) ports of $ABA$, respectively. Bottom panel: The corresponding simulated results.

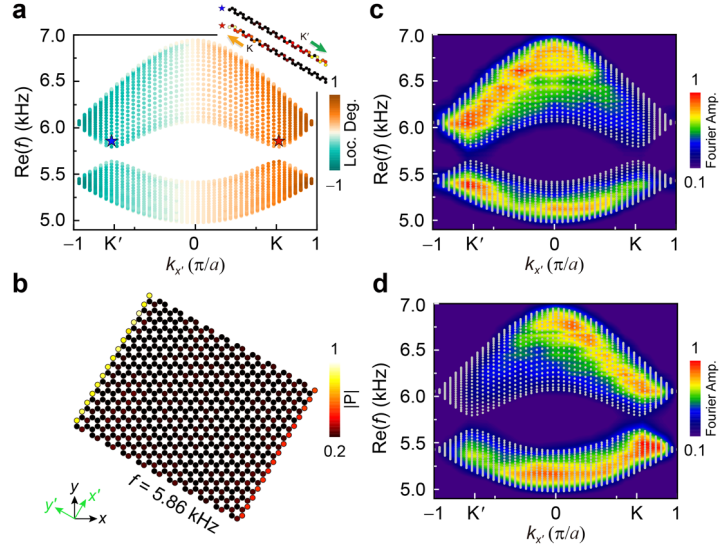

**Fig. 3 | Observation of valley-dependent skin effect. a**, Real part of the projected dispersions of a ribbon, which is periodic in the $x'$ direction and finite in the $y'$ direction. The positive and negative values of the colors indicate the localization degree of the bulk states at the upper-left and lower-right boundaries, respectively. Inset: Field distributions of two eigenstates denoted by blue and red stars in the dispersions. **b**, Measured pressure field distribution at 5.86 kHz. It is mainly localized on the upper-left and lower-right boundaries, visualizing the skin modes. **c**, Measured (bright color) projected dispersions by Fourier transforming the pressure field at the lower-right boundary. White dots represent the simulated results. **d**, The same to **c**, but for the upper-left boundary.

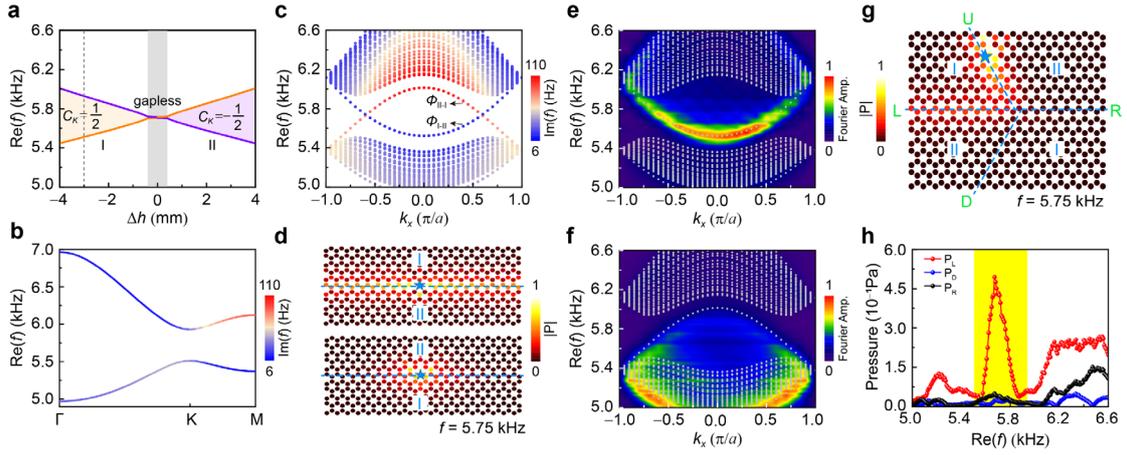

**Fig. 4 | Valley-projected edge states with boundary-dependent lifetimes and their anomalous beam splitting. a**, Phase diagram revealed by the real part of the band-edge frequencies at $K$ point as the function of $\Delta h = h_2 - h_1$, where $(h_1 + h_2)/2 = 30$ mm. The gray shadow ($|\Delta h| < 0.4$ mm) corresponds to bulk band gap closure. The light orange ($\Delta h < -0.4$ mm) and light purple ($\Delta h > 0.4$ mm) areas with opposite valley Chern numbers are denoted as phase I and II, respectively. **b**, Bulk band dispersions of phase I with $\Delta h = -3$ mm (gray dashed line in **a**). **c**, Complex projected dispersions for heterostructure I-II-I. The colors denote the imaginary part of the dispersions. **d**, Simulated pressure field distributions at 5.75 kHz for interface structures I-II (top) and II-I (bottom). **e**, **f**, Measured (bright color) dispersions by Fourier transforming the fields at the interfaces for structures I-II and II-I, respectively. White dots represent the simulated results. **g**, Simulated pressure field distribution at 5.75 kHz in the sample of an arrow-shaped interface intersection. **h**, Measured transmitted pressure at the port L, D, and R. The blue stars and lines in **d** and **g** represent the sound sources and the interfaces, respectively.